\documentclass[journal]{IEEEtran}

\usepackage{cite}
\usepackage[T1]{fontenc}
\usepackage[utf8]{inputenc}
\usepackage{amsmath,amssymb,amsfonts}
\usepackage{graphicx}
\usepackage{booktabs}
\usepackage{siunitx}
\usepackage{xcolor}
\usepackage[expansion=false]{microtype}
\usepackage{enumitem}
\usepackage{url}
\usepackage[hidelinks,breaklinks]{hyperref}

\DeclareSIUnit{\voltampere}{VA}
\DeclareSIUnit{\var}{var}

\title{The GIST Korea Test System: A Public-Data\\
Synthetic Model of the Korean Power Grid}

\author{Yun-Su~Kim%
\thanks{Y.-S. Kim is with the Department of Electrical Engineering and Computer
Science and the Research Institute for Solar and Sustainable Energies, Gwangju
Institute of Science and Technology (GIST), Gwangju, Republic of Korea.}}

\begin{document}
\maketitle

\begin{abstract}
No model of the Korean transmission system at native resolution is publicly
available, hindering reproducible research on one of the world's most
distinctive grids---an islanded interconnection with extreme
separation between generation and the Seoul Metropolitan Area load center and
heavy reliance on extra-high-voltage transmission. Working strictly from
public data, we present the \emph{GIST Korea test system}, a geographically
grounded synthetic model of the Korean grid. Unlike fully synthetic cases,
whose lines match no real corridor, and aggregated public Korean models, it
derives its 345 and 154\,kV layout from the OpenStreetMap power layer by a
multi-source shortest-path reassembly of overhead-line and underground-cable
geometry, gap-fills unmapped substations by a geographic minimum spanning
tree, and calibrates aggregate circuit length to
published statistics (94/100/104\% at 765/345/154\,kV). The current release
spans \num{2265} buses, \num{614} generation and renewable sources
(\SI{151}{\giga\watt}), \num{3185} AC circuits with explicit underground
sections, four HVDC converter links, \num{3413} transformers, and reactive
resources, in a PSS/E-compatible CSV schema with zero-sequence
and grounding data for unbalanced-fault studies. The model is distributed as a
\emph{frozen} operating point---taps, setpoints, and bus voltages settled once
offline---so a single deterministic Newton--Raphson pass reproduces a
\SI{91}{\giga\watt} evening-peak snapshot anchored to the observed 2025 summer
peak (\SI{2.3}{\percent} losses, no overloads), consistent with the public
KPG-193 model.
Standard-parameter dynamic, protection, and unit-commitment layers, and a
library of 24 hourly operating points derived by security-constrained unit
commitment, extend the dataset beyond a single power flow case. The dataset,
maps, and tooling are released as a citable, continuously maintained platform
for power flow, planning, and decarbonization studies.
\end{abstract}


\section{Introduction}
\label{sec:intro}

Realistic, publicly available grid models are the foundation of reproducible
research in power systems. For most national systems, however, the operational
network model is held by the system operator as confidential critical energy
infrastructure information (CEII) and is not publicly distributed. The Republic
of Korea is an acute example: no full-resolution model of the national
transmission network is publicly available to researchers, so methods are hard
to test on a realistic Korean case and published results on the Korean grid are
difficult to reproduce or compare. This motivates a model assembled, for research
purposes, entirely from publicly available data.

This scarcity is unfortunate because the Korean grid is structurally unusual and
therefore valuable as a test case. It is an \emph{isolated} interconnection with
no synchronous AC ties to neighboring countries; it exhibits a pronounced
spatial mismatch between bulk generation (nuclear and coal clustered on the
southern and eastern coasts) and demand (concentrated in the northwest in the
Seoul Metropolitan Area---Seoul, Incheon, and Gyeonggi); renewable penetration remains
comparatively low; and the bulk system leans heavily on a 765/345\,kV
extra-high-voltage (EHV) backbone with a dense 154\,kV sub-transmission mesh, of
which the urban portion is largely undergrounded.
These features---long-distance bulk transfer into
a single dominant load pocket, reactive-support-limited voltage profiles, and
islands fed by high-voltage direct current (HVDC)---are exactly the conditions under which planning and stability
methods are stressed.

\paragraph{The synthetic-grid landscape}
The community has responded to CEII restrictions with \emph{synthetic} grids that
are statistically and functionally similar to real systems while containing no
confidential data. The most influential line of work, the ACTIVSg family of
cases on the geographic footprint of Texas and the Eastern
Interconnection~\cite{birchfield2017structural,gegner2016methodology,birchfield2017validation,birchfield2018convergence,xu2017economic},
designs an \emph{entirely fictitious} transmission network whose statistical
characteristics (degree distributions, line-length distributions, loading
patterns) match validation criteria derived from real grids, but whose lines
correspond to no actual corridor. A complementary thread builds combined
transmission--distribution synthetic systems at very large scale~\cite{mateo2024td}.
For Korea specifically, the KPG-193 test system~\cite{song2024kpg193} is a
synthetic model developed from public data for decarbonization studies; to remain
computationally tractable and representative it applies clustering to obtain a
\num{193}-bus reduced model (with \num{407} transmission lines) in MATPOWER
format, validated through unit commitment and AC optimal power flow.

\paragraph{A different point in the design space}
The GIST Korea test system (named for the Gwangju Institute of Science and
Technology) occupies a region between these approaches. Like
ACTIVSg and KPG-193 it uses \emph{only} public information and contains no
confidential data. Unlike the fully synthetic ACTIVSg cases---whose lines
correspond to no actual corridor---it does not invent the transmission graph: it
recovers the \emph{actual} substation-to-substation circuits, routes, lengths,
circuit multiplicities, and overhead/underground character by tracing the
crowd-sourced OpenStreetMap (OSM)/OpenInfraMap power layer, preserving real
geographic coordinates throughout.
KPG-193 likewise draws its lines from OSM, so it too is grounded in real geography;
the difference is one of resolution. KPG-193 clusters the network to \num{193}
buses for market and expansion studies, which yields an aggregated backbone
rather than a native substation-level grid---in particular, a 193-bus reduction
cannot resolve the dense 154\,kV sub-transmission layer. The GIST system instead
keeps the topology \emph{un-clustered}: it resolves individual EHV substations,
the full 154\,kV sub-transmission and 22.9\,kV distribution layers, generator
step-up structure, and per-bank transformers, yielding an order of magnitude more
buses (\num{2265} in the current release vs.\ \num{193}). The result is a
high-resolution, geographically explicit, publicly documented model suitable for
AC power flow, contingency, and reactive-planning studies of the Korean system.

\paragraph{Contributions}
This paper makes the following contributions.
\begin{enumerate}[leftmargin=1.3em,itemsep=0.15ex,topsep=0.3ex]
\item A reproducible, public-data-only methodology for building a national-scale
grid model that derives its EHV topology from the OSM power layer via a
multi-source shortest-path reassembly of fragmented line geometry, with
feeder-based circuit-count estimation, explicit per-section overhead/underground
modeling, and calibration against published circuit-length statistics
(Sec.~\ref{sec:topology}).
\item A complete component model---buses, an EPSIS-registry-exact generation
fleet with step-up transformers and remote voltage control, renewables placed at
municipal resolution, regionally allocated load, two- and three-winding
transformers sized by standard bank ratings and an $N$$-$$1$ firm-capacity rule,
targeted reactive compensation, and flexible AC transmission systems
(FACTS)---serialized to a CSV schema compatible with PSS/E conventions that also
carries zero-sequence impedances, winding connections, and neutral grounding for
unbalanced-fault analysis (Sec.~\ref{sec:components}).
\item Companion data layers for dynamic simulation, protection, and production
costing, populated from published \emph{standard} parameters (PSS/E-library and
WECC generic models, IEEE/IEC protection guides, published fuel costs) rather
than confidential settings (Sec.~\ref{sec:layers}).
\item A general-purpose Newton--Raphson power flow solver with reactive limit
enforcement, explicit LCC/VSC converter models, and zero-impedance
regularization, plus an offline operating-point freezing step that bakes a
solved snapshot into the data so the model reproduces it in a single
deterministic pass (Sec.~\ref{sec:solver}), and a security-constrained
unit-commitment pipeline that derives a library of 24 hourly operating points
for a reference peak day (Sec.~\ref{sec:oplib}).
\item An evaluated evening-peak operating point anchored to the observed 2025
summer peak, a set of consistency checks against public benchmarks (published
circuit length, substation and transformer inventories, the public KPG-193
model, regional supply/demand balance), and an honest account of structural
limitations (Secs.~\ref{sec:results}--\ref{sec:limitations}).
\end{enumerate}

A central methodological principle throughout is that no confidential or
proprietary network data is used at any stage: every component count and
multiplicity is fixed solely by public facility inventories and conventional
transmission-design rules. Counts are deliberately \emph{not} fitted to a target:
the model is not steered toward matching the size of any non-public model;
quantities emerge from the public inventory and from physics. The model is under
continuous development, so component counts (including the number of buses)
evolve between releases; this paper describes release v5.2.0 (July 2026), and
the number of buses is therefore quoted in the text rather than fixed in the
system's name.

\section{Related Work}
\label{sec:related}

\paragraph{Statistically representative synthetic grids}
Birchfield \emph{et al.}~\cite{birchfield2017structural} established
\emph{structural characteristics}---substation count, transmission line-length
distributions, transformer ratios, and graph connectivity---as validation
criteria for synthetic networks, and Gegner \emph{et al.}~\cite{gegner2016methodology}
gave a methodology for laying out geographically realistic synthetic power flow
models. Realism metrics were further formalized
in~\cite{birchfield2017validation}, and reactive-power planning to obtain
solvable large synthetic cases was addressed
in~\cite{birchfield2018convergence}. These cases (e.g., ACTIVSg2000,
ACTIVSg10k) are deliberately disconnected from any real corridor. Our work shares
their goal of a CEII-free, publicly distributable model and their use of public
statistics as calibration targets, but differs in deliberately preserving real
geography and real EHV connectivity rather than generating a fictitious graph.

\paragraph{Korean synthetic models}
KPG-193~\cite{song2024kpg193} is, to our knowledge, the most directly related
prior public Korean model. It likewise extracts lines from OSM via the Overpass API
and assigns parameters from manufacturer catalogs, but it clusters the system to
\num{193} buses for tractability in market/decarbonization optimization and
adopts the MATPOWER format. The GIST Korea test system is complementary: it targets
AC power flow and planning fidelity at native resolution (EHV substations, a
distribution layer, and generator step-up structure), follows PSS/E-style data
conventions, and is exercised by AC power flow convergence rather than (or in
addition to) market simulation. Because it resolves an explicit 154/22.9\,kV
distribution interface, the GIST model is also a natural transmission-side anchor
for combined transmission--distribution coupling studies~\cite{mateo2024td}. The
two models can be used together---e.g.,
KPG-193 for fast economic studies and the GIST model for detailed network
analysis---and we cross-check the solved operating point against KPG-193 in
Sec.~\ref{sec:results}.

\paragraph{Tooling}
We build on \texttt{pandapower}~\cite{thurner2018pandapower}, an open Python
library for modeling and steady-state analysis, and adopt conventions consistent
with MATPOWER~\cite{zimmerman2011matpower}. We normalize to the data conventions
of PSS/E (Power System Simulator for Engineering; the industry-standard power flow
package from Siemens PTI, Power Technologies International), documented in its
program operation manual~\cite{psse2013}, and the
underlying power flow theory follows standard
references~\cite{grainger1994,kundur1994}. The unit-commitment pipeline of
Sec.~\ref{sec:oplib} uses the open-source HiGHS mixed-integer
solver~\cite{huangfu2018}, and the standard dynamic models of
Sec.~\ref{sec:layers} were cross-checked against the open-source ANDES
simulator~\cite{cui2021andes}.

\section{Design Principles and Data Sources}
\label{sec:principles}

\subsection{Principles}
\label{sec:design}
The model follows three rules that make it reproducible and keep it free of
confidential information.
\begin{enumerate}[leftmargin=1.3em,itemsep=0.15ex,topsep=0.3ex]
\item \textbf{Public data only.} Every element derives from publicly available
sources (Sec.~\ref{sec:sources}). No proprietary or operator-internal network
model is used at any stage.
\item \textbf{Bottom-up component counts.} The numbers of buses, lines,
transformers, and circuits follow from public facility inventories and
conventional transmission-design rules (e.g., $N$$-$$1$ firm transformer banks and
typical double-circuit corridors); they are not tuned to reproduce the size of any
non-public model.
\item \textbf{Aggregate calibration only.} Where public statistics exist (e.g.,
total transmission circuit-km by voltage level from the Electric Power Statistics
Information System, EPSIS), we use them as system-wide calibration targets, never
to fit any individual circuit.
\end{enumerate}
Two corollaries are applied consistently. Only facilities \emph{in operation}
at the anchor date are modeled: planned reinforcements from the national
transmission plans are deliberately excluded, even where their absence leaves a
documented weakness in the solved state. And every synthetic element whose
position (rather than existence) is approximate is flagged by a naming prefix so
that measured and estimated data remain distinguishable. The result is a
synthetic model whose layout is grounded in real, public geography wherever OSM
coverage allows, and is filled in by documented heuristics elsewhere;
Sec.~\ref{sec:limitations} states explicitly which parts are measured and which
are estimated.

\subsection{Public data sources}
\label{sec:sources}
The model is assembled from the following public sources:
\begin{itemize}[leftmargin=1.2em,itemsep=0.15ex,topsep=0.3ex]
\item \textbf{OpenStreetMap / OpenInfraMap}~\cite{osm} via the Overpass API ---
geolocated substations and transmission-line geometry (\texttt{power=line|cable},
voltage tags) for the 765/345/154\,kV network, including the overhead/underground
distinction.
\item \textbf{EPSIS}, operated by the Korea Power Exchange
(KPX)~\cite{epsis} --- aggregate transmission circuit length, substation and
transformer statistics, generation by fuel, monthly settlement fuel costs, the
registered-generator inventory (per-unit ratings and commissioning/retirement
status), and regional figures used as \emph{calibration targets}.
\item \textbf{Korea Energy Agency (KEA)} renewable statistics~\cite{kea} ---
municipal-level (si/gun/gu) solar and wind deployment.
\item \textbf{Regional electricity-demand statistics}~\cite{kesis} ---
province-level consumption shares for load allocation.
\item \textbf{Korea Electric Power Corporation (KEPCO) public
statistics}~\cite{kepcostat} and \textbf{manufacturer
disclosures} (Hyosung, ABB, LS) --- generator unit ratings, HVDC and FACTS
commissioning.
\item \textbf{Public one-line diagram} of the Jeju power
system~\cite{jeju_diagram} --- used \emph{visually} to cross-validate the Jeju
ring network and HVDC landing points (the mainland 345\,kV corridors are taken
from the OSM/OpenInfraMap layer above).
\item \textbf{KPG-193}~\cite{song2024kpg193} --- per-kilometer line parameters
(R/X/B) for standard Korean overhead conductors, since OSM does not carry
electrical parameters; underground-cable parameters follow published XLPE cable
data.
\item \textbf{Administrative boundaries}~\cite{kostat_admin} --- full-resolution
province polygons for the point-in-polygon regional assignment of
Sec.~\ref{sec:topology}.
\item \textbf{Published standards and model libraries} --- PSS/E-library dynamic
models~\cite{psse2013}, WECC generic renewable models~\cite{wecc2014},
IEEE~C37.102/C37.106 and IEC~60255 protection
guides~\cite{ieee_c37_102,ieee_c37_106,iec60255} for the standard-parameter data
layers of Sec.~\ref{sec:layers}.
\end{itemize}

\section{Network Topology Construction}
\label{sec:topology}

The transmission graph---which substations connect to which, along what route,
at how many circuits, and whether overhead or underground---is the most
consequential part of the model for power flow and the hardest to obtain from
public data. Throughout, a \emph{corridor}
denotes the physical substation-to-substation route (the transmission
right-of-way), whereas a \emph{circuit} is one three-phase line carried on that
route; a single corridor may carry several circuits (a double-circuit corridor
carries two). We build the graph in three stages, the last of which is adopted.

\subsection{Geographic primitives}
All inter-node distances use the Haversine great-circle formula \cite{sinnott1984}: for two points
$i$ and $j$ with latitudes $\phi_i,\phi_j$ and longitudes $\lambda_i,\lambda_j$,
writing $\Delta\phi=\phi_j-\phi_i$ and $\Delta\lambda=\lambda_j-\lambda_i$, the
great-circle distance is
\begin{equation}
d_{ij}=2R_E\arcsin\sqrt{\sin^{2}\frac{\Delta\phi}{2}
+\cos\phi_i\cos\phi_j\sin^{2}\frac{\Delta\lambda}{2}},
\end{equation}
where $R_E=\SI{6371}{\kilo\meter}$ is the Earth's mean radius. Where a route is
not traced from mapped geometry, physical line length is obtained from this
straight-line distance times a routing/detour factor of $1.15$---the median
ratio of traced route length to straight-line distance over all mapped overhead
circuits in the model ($1.1499$, $n\approx\num{1950}$). Each bus is assigned to
its administrative province by a \emph{point-in-polygon} test against the
full-resolution official province boundaries~\cite{kostat_admin} (buses on
reclaimed land, offshore platforms, or islands that fall outside every polygon
are assigned to the nearest province; the largest such offset is
\SI{1.8}{\kilo\meter}), and to one of six operating regions
(Seoul Metropolitan Area, Gyeongsang, Chungcheong, Jeolla, Gangwon, Jeju).
For naming, each coordinate also inherits its nearest city
$c(i)=\arg\min_{j\in\mathcal{C}} d_{ij}$ over the city set $\mathcal{C}$---the
one-nearest-neighbor rule \cite{cover1967}. (Earlier releases used the
nearest-city rule for the province assignment as well; replacing it with the
polygon test re-assigned 244 buses near provincial borders and, with the
load-conservation rule of Sec.~\ref{sec:load}, left every provincial demand
share unchanged.)

\subsection{Stage 1 -- nearest-neighbor mesh (deprecated)}
An initial coordinate-nearest mesh over-produced corridors (345\,kV at
\SI{174}{\percent} and 154\,kV at \SI{132}{\percent} of published route length)
and created many implausible $>$\SI{50}{\kilo\meter} edges, motivating a more
principled construction.

\subsection{Stage 2 -- geographic minimum-spanning-tree backbone (fallback)}
A minimum spanning tree (MST) of a connected, edge-weighted graph is the subset
of edges that connects every vertex with the smallest possible total edge weight
while containing no cycles. Taking substations as vertices and the Haversine
distance $d_{ij}$ as edge weight, the MST is
\begin{equation}
T^{\star}=\arg\min_{T\in\mathcal{T}}\;\sum_{(i,j)\in T} d_{ij},
\end{equation}
where $\mathcal{T}$ is the set of spanning trees of the complete substation
graph: the shortest-total-length network that still links every
substation---a reasonable first approximation to a length-minimizing
transmission backbone when measured routes are unavailable. We compute it with
Prim's algorithm \cite{prim1957}, which grows a tree from a root vertex by
repeatedly adding the lowest-weight edge that reaches an as-yet-unconnected vertex
until all are included; a multi-source minimum spanning \emph{forest} generalizes
this to several roots, producing one tree per root.

In this construction the 345\,kV backbone is a single tree (its minimum spanning
tree), whereas the 154\,kV layer is a spanning forest---one tree rooted at each
345/154\,kV injection point. Short distance-gated loop edges are then added for
$N$$-$$1$ redundancy, and circuit counts are assigned by corridor length (short
corridors double-circuit, long corridors single-circuit). This backbone was
reconciled with a list of real 345\,kV trunk
corridors visually transcribed from the public OpenInfraMap rendering of the OSM power
layer~\cite{osm} (21 corridors added; 16 non-existent links pruned). Because an
MST is purely radial and minimizes total length rather than reproducing actual
routing or meshing, this MST$+$corridor construction is retained only as a code
fallback and is \emph{not} the default.

\subsection{Stage 3 -- real OSM line topology (adopted)}
\label{sec:osm}
OpenInfraMap renders the OSM power layer, in which the Korean transmission system
is broadly mapped. We query the live Overpass API for 765/345/154\,kV lines and
compare the measured geometry length against published circuit length: coverage
is \SI{100}{\percent}, \SI{96}{\percent}, and \SI{79}{\percent} at
765/345/154\,kV respectively. The lines are then reassembled into
substation-to-substation circuits.

\paragraph{Multi-source shortest-path reassembly}
In OSM, each overhead line is fragmented into many short \texttt{way} segments
(split at towers), so a corridor is not a single object. We seed every
substation's region with the line nodes inside a tight radius of its centroid and
grow the regions by multi-source shortest path (Dijkstra's algorithm
\cite{dijkstra1959} run from all seeds simultaneously), assigning each line node
$v$ to the substation $a(v)=\arg\min_{s\in\mathcal{S}}\operatorname{dist}(s,v)$,
where $\mathcal{S}$ is the seed set and $\operatorname{dist}(s,v)$ the
shortest-path length along the line graph---a geodesic Voronoi partition of the
line graph. Where two substation regions meet, we recover a direct circuit between
them whose length is the tower-following shortest path. Large urban substations
whose feeders end outside the tight radius would otherwise be dropped; a
\emph{snapping recovery} step assigns the nearest line node as a seed to any
substation left without a region, without letting it capture nodes belonging to
another substation's region.

\paragraph{Feeder-based circuit counting}
Circuit multiplicity per corridor is estimated from the circuit tag of the
\emph{first} segment leaving the substation (the feeder). A naive
maximum-along-path estimate inflates length because a high-circuit line crossing
the middle of a long corridor propagates its tag along the whole path
(345\,kV at \SI{114}{\percent}); the feeder-based estimate is robust
(\SI{106}{\percent}).

\paragraph{Adopted topology and cross-validation}
At 345\,kV we obtain \num{117} real OSM circuits and gap-fill only the \num{19}
substations OSM does not reach with the geographic MST. The recovered corridors
were cross-checked against the visually transcribed corridor list from Stage~2: the automated
OSM trace confirmed 8 corridors, validated 7 pruned (non-existent) links, and
revealed one link the manual prune had removed in error, after which OSM was
adopted as the primary source. At 154\,kV we obtain
\num{689} real OSM circuits and gap-fill \num{130} unreached substations;
geographic plausibility was confirmed on mountainous and island spans (e.g.,
Bonghwa--Uljin \SI{62}{\kilo\meter}; Sinan island chains). The 765\,kV network (8
substations) is small and already accurate from public corridors, so it is
retained. Three island switching stations carry no distribution load and are
modeled as zero-load. Since the initial release the adopted topology has been
refined circuit-by-circuit against the OSM rendering and satellite imagery
(corrected circuit multiplicities, re-measured spans, and pruned artifacts);
these revisions are documented in the released changelog and are reflected in
the calibration figures below.

\paragraph{Underground cables and mixed circuits}
\label{sec:cables}
Urban Korean sub-transmission is largely underground, and treating those
circuits as overhead misstates both their series impedance and---because
cross-linked-polyethylene (XLPE) cable charging is an order of magnitude larger
per kilometer---the reactive balance of the metropolitan networks. The OSM
power layer distinguishes \texttt{power=line} from \texttt{power=cable}, and
each circuit in the dataset therefore carries, alongside its total route length
$L$, an explicit underground length $\ell_c$ ($0\le\ell_c\le L$). Electrical
parameters of a mixed circuit are composed \emph{per section} in series: with
overhead and cable per-kilometer positive-sequence parameters
$(r,x,b)_{\mathrm{oh}}$ and $(r,x,b)_{\mathrm{cb}}$,
\begin{equation}
\label{eq:mix}
\begin{aligned}
r_1 &= r_{\mathrm{oh}}(L-\ell_c) + r_{\mathrm{cb}}\,\ell_c, &
x_1 &= x_{\mathrm{oh}}(L-\ell_c) + x_{\mathrm{cb}}\,\ell_c,\\
b_1 &= b_{\mathrm{oh}}(L-\ell_c) + b_{\mathrm{cb}}\,\ell_c, &&
\end{aligned}
\end{equation}
i.e., series impedances add and shunt charging adds along the route. The
zero-sequence parameters compose by the same rule with per-section conventions
$x_0=3x_1$, $b_0=0.6\,b_1$ on overhead sections (ground-return dominated) and
$x_0=x_1$, $b_0=b_1$ on cable sections (metallic sheath return), with
$r_0=3r_1$ throughout (Sec.~\ref{sec:sequence}). Ratings follow the standard
Korean conductor data---overhead 345\,kV ACSR 480\,mm$^2\times$4-bundle
(\SI{2173}{\mega\voltampere} per circuit), overhead 154\,kV ACSR
410\,mm$^2\times$2-bundle (\SI{452}{\mega\voltampere}), XLPE 2500\,mm$^2$
(\SI{1023}{\mega\voltampere}) and XLPE 2000\,mm$^2$
(\SI{430}{\mega\voltampere}) underground---and a mixed circuit takes the rating
of its \emph{weakest series section} (the cable), since sections in series carry
the same current. Short-term emergency ratings are set at 120\% and 125\% of
the continuous rating following Korean planning practice. The current release
carries \num{817} circuits with underground sections (\num{682} fully
underground), totaling $\approx$\num{2372} cable circuit-km, concentrated in
the Seoul, Incheon, and Busan metropolitan 154\,kV networks and a small number
of named 345\,kV cable tunnels traced from OSM.

\paragraph{Synthetic substation gap-filling}
EPSIS lists \num{788} 154\,kV substations nationwide, but OpenInfraMap maps only
about \num{697} (\SI{88}{\percent}); the operator withholds the names and
coordinates of the rest, mostly urban indoor and underground gas-insulated
stations. The missing stations exist but are unlocated, and the gap is
not benign: spreading a province's demand over only the mapped subset inflates the
load per substation (Busan averages \SI{283}{\mega\watt} per station against a
national mean of $\approx$\SI{110}{\mega\watt}) and so manufactures artificial 154\,kV
overloads. We treat this as \emph{count-constrained spatial disaggregation}---the
aggregate counts are trustworthy but the individual positions are not, so synthetic
nodes are added to the observed set to meet the known regional totals. Three
principles govern the construction: regional counts are fixed by the public
statistic rather than invented; only individual positions are approximated, spread
as evenly as possible without clustering or alignment; and every synthetic node is
flagged (a \texttt{syn} prefix) to stay distinct from measured data.

The deterministic procedure fills, for each administrative region $r$, a deficit
\begin{equation}
d_r=\max\!\big(0,\; n^{\mathrm{EPSIS}}_r-n^{\mathrm{OSM}}_r\big),
\end{equation}
allocating the $d_r$ nodes to cities in proportion to their mapped-substation
density. Each node is then placed around its anchor city by a Vogel (sunflower)
phyllotaxis spiral: the $i$-th node is offset at angle
\begin{equation}
\theta_i = i\,\theta_g,\qquad \theta_g = 137.508^\circ\ \text{(the golden angle)},
\end{equation}
and radius $r_i\approx\SIrange{2.8}{5.5}{\kilo\meter}$ (longitude scaled by
$\cos\phi$). Because the golden angle is an irrational fraction of a full turn, the
points never fall on any line or lattice and fill the plane as evenly as
possible---the principle behind seed packing in a sunflower head---so several nodes
around one city disperse without overlap, and, using no random numbers, the layout
is fully reproducible. Each synthetic station contributes one 154\,kV bus and one
22.9\,kV distribution bus and is joined to the nearest real 154\,kV bus by the same
gap-fill spanning forest used above; Jeju is excluded, since its fixed public ring
would leave synthetic nodes stranded. This brings the modeled 154\,kV
distribution-station count to \num{799} (\SI{101}{\percent} of the EPSIS total)
and restores a realistic load density, removing the artificial urban overloads.
Since the initial release, individual synthetic stations have been progressively
re-identified with real named substations from the OSM rendering (and deleted or
relocated where the public inventory demanded); about \num{90} remain synthetic
in the current release. The redistribution that respects a per-station capacity
limit is described in Sec.~\ref{sec:load}.

A second, better-grounded class of inferred stations arises at 345\,kV: the OSM
345\,kV layer contains cable and line \emph{endpoints} with no mapped
substation---predominantly underground urban infeeds. A station is instantiated
at the measured endpoint position (an \texttt{inf} prefix records that its
identity, not its location, is inferred). Thirty-one such stations were
introduced initially; screening them against the EPSIS 345\,kV transformer
inventory and provincial demand showed the set to be over-complete, and the
current release retains eleven, several others having since been re-identified
with named substations. Each class of element thus carries an explicit evidence
grade: measured OSM elements (position and identity), endpoint-inferred stations
(measured position, inferred identity), and golden-angle synthetic stations
(inferred position and identity).

\paragraph{Circuit-length definition and calibration}
EPSIS publishes transmission length as \emph{circuit-km} (C-km), i.e., route
length times the number of three-phase circuits, not route-km. (A decisive check:
the real 765\,kV system has only $\sim$\SI{580}{\kilo\meter} of public route but
EPSIS reports \SI{1024}{\kilo\meter}, a ratio of $\sim$$1.8$ consistent with
circuit counting.) Calibrating the adopted topology to this definition yields
circuit-length ratios of 765/345/154\,kV $=94/100/104\%$ and an AC total of
$\approx$\SI{102}{\percent} (Fig.~\ref{fig:validation}a). Because the real OSM
lines are less densely looped than the synthetic mesh, both the line count and
the system losses fall relative to Stage 2.

\subsection{Jeju island network and HVDC}
\label{sec:jeju}
Because the geographic MST excludes Jeju (which is fed only by HVDC), Jeju's
154\,kV substations were initially isolated. Following a public Jeju one-line
diagram~\cite{jeju_diagram} (used visually only), we add the coastal ring network
(converter stations, Sin-Jeju, Dong-Jeju, and the southern and western nodes) and
eight named wind farms, after which all Jeju buses are connected and served.

The four HVDC links in the system are represented by explicit two-terminal
converter models rather than AC equivalents. Three line-commutated-converter (LCC)
links---the Jeju--Haenam and Jeju--Jindo submarine links and the mainland
Bukdangjin--Godeok link---are described in a \texttt{dcline.csv} file (converter
bridge count, firing/extinction angles, commutating reactance, and DC
resistance) at \emph{pole} resolution, one record per pole of each bipole; one
voltage-source-converter (VSC) link, Jeju--Wando (commissioned in late 2024), is
described in \texttt{vscdcline.csv}, with its mainland terminal operated in
AC-voltage-control mode so that the converter's reactive capability supports the
weak coastal network, as in actual operation. When these DC files are present the
loader supersedes the corresponding \texttt{HVDC\_*} AC-equivalent lines with the
converter models, so each corridor carries its scheduled firm power through a
physical converter representation and is excluded from the AC circuit-length
accounting; deleting the DC files reverts to the legacy AC-equivalent behavior.
Planned but not-yet-commissioned HVDC and 765\,kV reinforcements are deliberately
excluded.

\begin{figure}[htbp]
\centering
\includegraphics[width=\columnwidth]{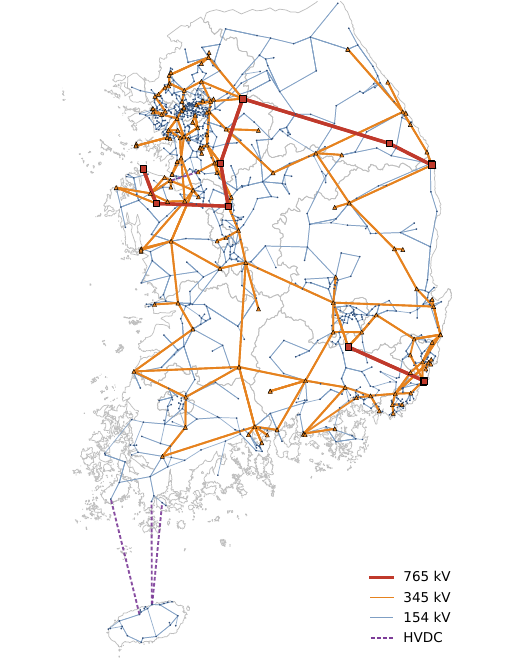}
\caption{Geographic topology of the GIST Korea test system (release v5.2.0).
Province boundaries from~\cite{kostat_admin}; interactive version
online~\cite{gist_map}.}
\label{fig:topology}
\end{figure}

Figure~\ref{fig:topology} shows the adopted geographic topology.

\section{Component Modeling}
\label{sec:components}

\subsection{Buses and identifier scheme}
Buses arise on five paths: voltage-level buses per substation; EHV tertiary
(23\,kV) windings; a 154/22.9\,kV distribution layer; generator switchyards; and
generator machine-terminal buses. Synthetic bus numbers are assigned in
voltage-banded ranges (765:\,\num{10000}; 345:\,\num{20000}; 154:\,\num{40000};
23:\,\num{60000}; 22.9:\,\num{70000}; machine:\,\num{80000}; switchyard:\,\num{90000}).
Table~\ref{tab:scale} summarizes the resulting scale; the bus voltage classes are
broken down in Table~\ref{tab:buses}.

\begin{table}[htbp]
\centering
\caption{Scale of the GIST Korea test system (release v5.2.0).}
\label{tab:scale}
\small
\begin{tabular}{@{}lr@{}}
\toprule
Element & Count \\
\midrule
Buses (PQ \num{2038} / PV \num{226} / REF 1) & \num{2265} \\
Generation \& renewable sources & \num{614} \\
\quad dispatchable units (nuclear/coal/gas/oil/hydro) & \num{233} \\
\quad distributed solar (photovoltaic) sites & \num{195} \\
\quad distributed wind sites & \num{44} \\
\quad non-central baseload sites (fuel cell, biomass, \ldots) & \num{142} \\
Loads & \num{887} \\
AC line circuits (with underground sections \num{817}) & \num{3185} \\
HVDC converter links (3 LCC / 1 VSC; LCC per pole) & 4 \\
Two-winding transformers & \num{3073} \\
Three-winding transformers & \num{340} \\
Shunt elements (switched \num{704} / fixed 8) & \num{712} \\
FACTS devices & \num{11} \\
\midrule
Installed generation capacity & \SI{150.6}{\giga\watt} \\
Operating regions & 6 \\
\bottomrule
\end{tabular}
\par\vspace{3pt}
{\scriptsize\raggedright Bus types: PQ (load), PV (voltage-controlled),
REF (reference/slack).\par}
\end{table}

\begin{table}[htbp]
\centering
\caption{Buses by voltage class (release v5.2.0).}
\label{tab:buses}
\small
\begin{tabular}{@{}lr@{\hskip 1.4em}lr@{}}
\toprule
Class & Buses & Class & Buses \\
\midrule
765\,kV & 13   & 22.9\,kV (distr.) & 799 \\
345\,kV & 151  & 18\,kV (machine)  & 148 \\
154\,kV & 962  & 23\,kV (tertiary) & 113 \\
        &      & other gen.\ ($\le$22\,kV) & 79 \\
\bottomrule
\end{tabular}
\end{table}

\subsection{Generation and dispatch}
The generation fleet comprises \num{233} dispatchable units at approximately one
hundred power plants, plus \num{381} distributed renewable and non-central
sites. Unit ratings are taken \emph{exactly} from the public EPSIS
registered-generator inventory (per-unit nameplate records), against which the
fleet was reconciled unit-by-unit: combined-cycle blocks carry their registered
gas-/steam-turbine composition, retired units are removed as of the anchor date
(with two nuclear units retained out of service pending license renewal), and
newly commissioned plants are included. Each conventional unit is placed on a
machine-terminal PV bus behind a generator step-up (GSU) transformer of
\SI{13}{\percent} impedance and controls its high-voltage switchyard by
\emph{remote} voltage control (Sec.~\ref{sec:remote}); the slack is a mid-size
coastal coal unit chosen so that the settled slack output lies inside its
governor band (Sec.~\ref{sec:solver}).

Renewables are placed at municipal resolution: the public plant-level registry
($\approx$\num{187000} records, \SI{29}{\giga\watt} of solar) is aggregated to
the country's 313 municipalities, geocoded to municipal centroids
(\SI{99.6}{\percent} match), and assigned to the nearest substation carrying a
distribution bus; wind (\SI{2.1}{\giga\watt}) combines named farms with
municipal aggregates. Non-central baseload---fuel cells, biomass, small hydro,
industrial cogeneration, and tidal, \SI{3.3}{\giga\watt} in
all---is placed the same way and treated as must-run injection; its inclusion is
what lets the model reproduce the published off-peak marginal price (a coal-set
overnight system marginal price) rather than an artificially high one.

\begin{figure}[htbp]
\centering
\includegraphics[width=\columnwidth]{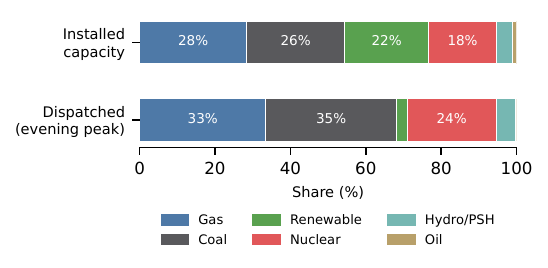}
\caption{Generation mix: installed capacity versus dispatch at the evening-peak
snapshot.}
\label{fig:genmix}
\end{figure}

Dispatch is set by merit order to meet load plus a small margin: renewables and
non-central units must-run first, then nuclear near capacity, then coal at a
reduced factor (reflecting environmental dispatch, seasonal particulate-matter
management, and retirement of aging units), then flexible liquefied natural gas
(LNG), hydro, and pumped storage proportionally, with LNG the largest installed
fuel. The base dispatch is rebalanced so that non-slack generation equals load
times one plus a loss margin---an invariant that keeps the settled slack output
small and physically meaningful. The resulting shares
(Fig.~\ref{fig:genmix}) are, by installed capacity,
gas\,$28$/coal\,$26$/renewable\,$22$/nuclear\,$18$\,\%, consistent with the
published Korean mix. Because the reference snapshot is anchored to the observed
system peak---which in Korea now falls \emph{after sunset}
(Sec.~\ref{sec:results})---the dispatched shares at the snapshot are
coal\,$35$/gas\,$33$/nuclear\,$24$\,\% with solar contributing nothing, a
deliberate representation of the evening net-load peak that stresses the
system's conventional fleet.

\subsection{Load and its regional allocation}
\label{sec:load}
\begin{figure}[htbp]
\centering
\includegraphics[width=\columnwidth]{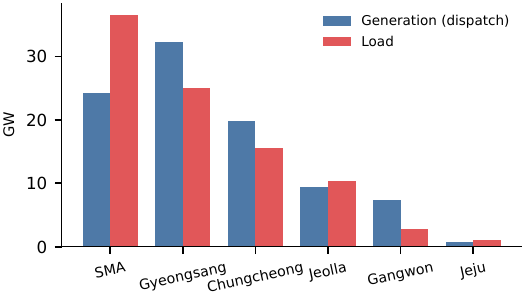}
\caption{Regional generation versus load at the evening-peak snapshot.}
\label{fig:regional}
\end{figure}

Regional demand totals are distributed to buses using deterministic weights that
are renormalized so that each province's sum exactly preserves its published
consumption share---an invariant maintained under every subsequent topology
revision (when buses are re-assigned between provinces or synthetic stations are
removed, the affected loads are re-scaled proportionally within each province so
that provincial totals are conserved; the summed absolute deviation of the 17
provincial shares from the published statistics is 1.9 percentage points).
Distribution-level (154\,kV) substations receive
a single aggregated load on their 22.9\,kV bus; 345\,kV substations receive load
on their 154\,kV bus; the load power factor is $0.95$. To keep any one station
within a physically plausible size, the per-substation load is capped at
\SI{216}{\mega\watt} (four \SI{60}{\mega\voltampere} banks at \num{0.9} power
factor, the standard four-bank 154\,kV limit), and any excess is redistributed to
under-loaded stations in the same province by water-filling---raising the lowest
stations first until no station exceeds the cap. Direct 154\,kV loads attached to
345/154\,kV substations carry a reduced weight, since only a few large customers
take supply at 154\,kV, and the single largest point load, the Samsung Pyeongtaek
campus at Godeok, is fixed at its publicly reported \SI{789}{\mega\watt} rather than
sampled. The total served load at the reference snapshot is
\SI{91.3}{\giga\watt}, anchored to the \SI{96}{\giga\watt} market-demand peak
recorded on 25 August 2025. The load records support the full ZIP
(constant-impedance/constant-current/constant-power) decomposition, with
published EMS-derived ZIP shares available as a documented option; the frozen
snapshot uses constant power. The defining structural feature of the Korean
system is visible in this allocation (Fig.~\ref{fig:regional}): the Seoul
Metropolitan Area consumes \SI{37}{\giga\watt} but dispatches only
$\approx$\SI{24}{\giga\watt} of local generation, importing the balance over the
EHV backbone, while Gyeongsang and Chungcheong are large net exporters.

\subsection{Transformers}
Three transformer populations are modeled: generator GSUs (two-winding);
154/22.9\,kV distribution main transformers sized by an $N$$-$$1$ \emph{firm}
capacity criterion (a bank outage leaves the remainder able to carry the full
load); and 765/345/23 and 345/154/23\,kV autotransformers (three-winding, with a
floating-delta 23\,kV tertiary). For power flow,
each three-winding unit is decomposed into its star-equivalent; on-load tap
changers on step-down transformers regulate their controlled bus to
$\approx$\SI{1.0}{pu} (Sec.~\ref{sec:oltc}).

The 345/154\,kV autotransformer population follows the Korean field standard of
\SI{500}{\mega\voltampere} banks, two to four banks per substation. The bank
count at each site is set by the same $N$$-$$1$ firm rule as the distribution
transformers,
\begin{equation}
n_{\mathrm{bank}} \;=\; \operatorname{clip}\!\Big(
\big\lceil S_{\mathrm{site}}/500 \big\rceil + 1,\; 2,\; 4\Big),
\end{equation}
where $S_{\mathrm{site}}$ is the through-flow (in \si{\mega\voltampere}) of the
site in the solved base case---so that the loss of one bank leaves the
remaining $n-1$ able to carry the base flow within rating. The resulting
national population (\num{316} banks, \SI{158}{\giga\voltampere}) matches the
EPSIS 345/154\,kV transformer-capacity statistic to within
\SI{3}{\percent}, with a 2/3/4-bank distribution consistent with KEPCO
practice. Impedances preserve the standard \SI{12}{\percent} per-bank impedance
\emph{on the bank's own rating}---when a bank rating changes, the system-base
per-unit impedance is rescaled so the own-base percentage is invariant. The
765/345\,kV interface uses \SI{2000}{\mega\voltampere} banks
(\SI{104}{\percent} of the EPSIS capacity). All transformer records carry a
winding-connection code (\texttt{ynd} for GSUs and distribution transformers,
\texttt{ynynd} for EHV autotransformers) and per-winding neutral-grounding
impedance fields, used by the sequence networks of Sec.~\ref{sec:sequence}.

\subsection{Sequence data for unbalanced faults}
\label{sec:sequence}
The dataset carries the zero-sequence system alongside the positive-sequence
one, so that single-line-to-ground, line-to-line, and double-line-to-ground
faults can be analyzed by the classical symmetrical-component construction: the
negative- and zero-sequence Th\'evenin impedances at the fault point are folded
into an effective fault impedance in the positive-sequence network
\cite{grainger1994}. Because zero-sequence impedances are not published
per-circuit, they are seeded by documented standard conventions---overhead
sections $x_0=3x_1$, $b_0=0.6\,b_1$; cable sections $x_0=x_1$, $b_0=b_1$;
$r_0=3r_1$; mixed circuits composed per section by \eqref{eq:mix}---and machine
negative-sequence reactances default to $x_2=(x''_d+x''_q)/2$. The seeded
system passes the standard self-consistency checks (e.g., the computed
ground-fault current of an ungrounded pocket equals its capacitive charging
current, and the sequence voltage ordering
$V^{\mathrm{3ph}}<V^{\mathrm{LLG}}<V^{\mathrm{LL}}<V^{\mathrm{SLG}}$ holds at
the fault point), and the values are explicitly flagged as typical-value seeds
rather than measured data.

\subsection{Shunt compensation and FACTS}
Reactive resources comprise 765\,kV fixed line reactors (sized by computed line
charging), targeted 345\,kV switched capacitors placed \emph{only} at the
weak-voltage buses identified by the power flow solution, and distribution
power-factor capacitors placed only at large load substations (smaller
substations absorb their power-factor correction into a near-unity load power
factor, keeping reactive physics neutral without an explicit shunt). Because
per-device reactive set points (Mvar) are not available in public data,
set points are chosen freely to match the operating snapshot, while
\emph{locations and counts} are taken from publicly documented installations,
and the \emph{aggregate} capacitor and reactor capacities are reconciled against
the EPSIS reactive-equipment statistics (99\% and 101\% of the published totals,
respectively). The model
includes \num{11} FACTS devices (Table~\ref{tab:facts}), each publicly documented
as in operation: six 345\,kV static synchronous compensators (STATCOMs), one
154\,kV unified power flow controller (UPFC), two 154\,kV STATCOMs, and two Jeju
synchronous condensers. FACTS are modeled in power flow as voltage-controlling
generators with zero real power and symmetric reactive limits.

\begin{table}[htbp]
\centering
\caption{FACTS devices (all publicly documented as in operation).}
\label{tab:facts}
\footnotesize
\begin{tabular}{@{}llrr@{}}
\toprule
Device & Type & kV & Rating (Mvar) \\
\midrule
Migeum, Sin-Buyeong & STATCOM & 345 & 100,\,150 \\
Sin-Yeongju, Sin-Chungju & STATCOM & 345 & 400,\,400 \\
Donghae & STATCOM & 345 & 150 \\
Godeok & MMC STATCOM & 345 & 400 \\
Gangjin & UPFC & 154 & 80 \\
Sin-Jeju, Halla & STATCOM & 154 & 50,\,50 \\
Buk-Jeju, Jeju & Syn.\ condenser & 154 & 55,\,50 \\
\bottomrule
\end{tabular}
\par\vspace{3pt}
{\scriptsize\raggedright MMC: modular multilevel converter.\par}
\end{table}

\section{Standard-Parameter Data Layers}
\label{sec:layers}

Beyond the power flow case, the dataset ships three companion layers that make
the model usable for time-domain, protection, and production-cost studies.
All three are populated from \emph{published standard parameters}---no
operator settings are used or implied---and each layer resides in its own
sub-directory so that its presence is optional and the base power flow case is
unaffected.

\paragraph{Dynamics}
Every synchronous machine carries a standard PSS/E-library model chain---GENROU
or GENSAL machines with IEEE-type excitation systems (EXST1, EXAC1, ESST4B),
turbine governors (IEEEG1, GGOV1, HYGOV, GAST), and PSS2A stabilizers on large
units---and every inverter-based resource carries the WECC second-generation
generic models (REGCA1 with REECA1/REECB1 and plant controller
REPCA1)~\cite{psse2013,wecc2014}. Parameters are the typical values published
with these model libraries, scaled by machine rating, and an under-frequency
load-shedding table follows Korean staged practice. The assembled models were
cross-checked against the open-source ANDES implementation of the same
standards~\cite{cui2021andes}. These are \emph{generic} dynamic data: they make
the case exercisable for transient-stability studies, but no claim is made that
individual plant tuning matches the field.

\paragraph{Protection}
A protection layer provides distance relays on both ends of every AC circuit
(mho zones at the conventional 80\%/125\%/220\% reaches with standard zone
timers), inverse-time overcurrent relays per IEC~60255~\cite{iec60255},
transformer overcurrent backup, generator protection (under/over-frequency,
under/over-voltage, out-of-step) per IEEE~C37.102 and
C37.106~\cite{ieee_c37_102,ieee_c37_106}, power-swing blocking, and automatic
reclosing on overhead circuits only. All settings are generated
deterministically from these public standards by a documented script; they
represent conventional practice, not utility setting files.

\paragraph{Production-cost economics}
A unit-commitment layer assigns each unit typical commitment parameters
(minimum up/down times, ramp rates, start-up costs) by fuel class, quadratic
cost curves anchored to the published EPSIS monthly settlement fuel costs,
configuration-based combined-cycle definitions matching the registered
gas-/steam-turbine composition, and the five Korean-market reserve products with
their substitution hierarchy. This layer drives the operating-point derivation
of Sec.~\ref{sec:oplib} and supports market-style studies (system marginal
price and locational marginal price decompositions) on the same network.

\section{Power Flow Formulation and Solver}
\label{sec:solver}

The model is solved with a general-purpose loader built on
\texttt{pandapower}~\cite{thurner2018pandapower}, which converts the CSV schema
into an internal network and runs the power flow described below.

\subsection{Per-unit system and network equations}
All quantities are normalized to a \SI{100}{\mega\voltampere} base. The nodal
relation is $\mathbf{I}=\mathbf{Y}\mathbf{V}$, with off-diagonal
$Y_{ik}=-y_{ik}$ (series admittance between $i$ and $k$) and diagonal
$Y_{ii}=\sum_k y_{ik}+\sum y^{\text{sh}}_i$ (series plus shunt/charging/excitation
terms). The injected complex power at bus $i$ is
$S_i=V_i\sum_k (Y_{ik}V_k)^{\!*}$, giving the real power flow equations
\begin{align}
P_i &= V_i \sum_{k} V_k\,(G_{ik}\cos\theta_{ik}+B_{ik}\sin\theta_{ik}),\\
Q_i &= V_i \sum_{k} V_k\,(G_{ik}\sin\theta_{ik}-B_{ik}\cos\theta_{ik}),
\end{align}
with $\theta_{ik}=\theta_i-\theta_k$.

\subsection{Newton--Raphson solution}
With state $\mathbf{x}=[\boldsymbol{\theta},\mathbf{V}]^{\!\top}$ and mismatches
$\Delta P,\Delta Q$, each Newton--Raphson iteration \cite{tinney1967} solves the
sparse linear system
\begin{equation}
\begin{bmatrix}\Delta P\\ \Delta Q\end{bmatrix}
=\begin{bmatrix}\mathbf{H}&\mathbf{N}\\ \mathbf{M}&\mathbf{L}\end{bmatrix}
\begin{bmatrix}\Delta\boldsymbol{\theta}\\ \Delta\mathbf{V}/\mathbf{V}\end{bmatrix},
\end{equation}
and updates $\mathbf{x}$. PV rows/columns ($\Delta Q$, $\Delta V$) and the slack
rows/columns are removed, leaving $2N_{pq}+N_{pv}$ unknowns. Convergence is
declared on the infinity norm of the power mismatch
($\le$\SI{e-5}{pu}); the method exhibits second-order convergence (typically
3--6 iterations), with DC and flat-start fallbacks.

\subsection{Reactive limit enforcement}
A PV generator whose computed $Q$ violates $[Q_{\min},Q_{\max}]$ is switched to PQ
at the violated bound and its voltage freed (and restored to PV if its voltage
recovers). This models automatic voltage regulator (AVR)/excitation saturation
realistically and is essential to obtaining a credible peak operating point.

\subsection{HVDC converter settling and DC islands}
The LCC links are solved by alternating the AC power flow with the converter
equations (tap and firing/extinction angle within their bands, filter reactive
injection at the terminals) until the scheduled DC power and terminal conditions
are consistent, and the VSC link enforces its terminal control modes
(DC power/voltage on one side, AC voltage on the other) with converter losses.
Because Jeju is connected only through DC, it is an asynchronous island: the
loader detects DC-only islands and promotes a designated Jeju unit to island
swing automatically.

\subsection{Remote voltage-control outer loop}
\label{sec:remote}
In PSS/E a generator may regulate a \emph{remote} bus (the \texttt{reg\_bus}
field)---for example, a unit on an \SI{18}{\kilo\volt} machine terminal holding
the high-voltage switchyard beyond its GSU. Since the \texttt{pandapower}
generator regulates only its own bus, we wrap the
Newton--Raphson solve in an outer loop that adjusts each generator's local voltage
set point $u$ so the controlled bus reaches its target. Because the sensitivity
of the controlled-bus voltage to $u$ is unknown (it depends on system strength and
GSU impedance), the gain is set from a secant (finite-difference) estimate of the
inverse sensitivity $\Delta u/\Delta V_{\text{ctrl}}$. At outer iteration
$\nu$,
\begin{align}
e &\leftarrow V_{\text{set}} - V_{\text{ctrl}}, \\
g &\leftarrow \mathrm{clip}\!\left(\Delta u/\Delta V_{\text{ctrl}},\,0.3,\,5\right),\\
u &\leftarrow \mathrm{clip}\!\left(u+g\,e,\,0.85,\,1.20\right),
\end{align}
followed by a warm-started re-solve; the set point clip stands in for the
excitation limit. A generator is ``settled'' when its controlled bus reaches
target or it saturates against the set point limit; the loop terminates when all
remote generators settle. Modeling the GSU step-down without remote control would
artificially depress the high-voltage side, so this loop is decisive for accuracy
in machine-terminal$+$GSU models such as ours.

\subsection{OLTC and switched-shunt fixed-point control}
\label{sec:oltc}
Tap changers and switched shunts are, by default, frozen at their snapshot values
(to reproduce a given operating point). An optional active mode automatically
identifies controllable devices from each transformer's controlled-bus and
winding voltages---step-down two-winding units regulating a low-voltage
(distribution) bus, and EHV autotransformers regulating a 154\,kV bus---and wraps
the solve in a fixed-point iteration. For a regulated bus with target $V^{*}$ and
tap step $\delta$ ($=1\%$), the tap position is updated by
$\Delta\text{tap}=\mathrm{clip}\!\big(\mathrm{round}((V-V^{*})/\delta),-1,+1\big)$
per iteration, with a deadband to suppress $\pm1$-step hunting, while switched
capacitors/reactors are toggled on a hysteresis band; each update is followed by a
warm-started re-solve. Robustness for thousands of coupled OLTCs is obtained by
rolling back a diverging iteration to the last converged state and by terminating
at the minimum-error state once the maximum regulated-bus error plateaus. The two
modes answer different questions---``reproduce this snapshot'' versus ``where do
OLTCs settle''---so the active mode is off by default.

\subsection{Operating-point freezing and single-pass convergence}
\label{sec:freeze}
Under the \SI{91}{\giga\watt} peak snapshot the network runs close to its
reactive capability, so the solved state depends on several discrete control
loops---on-load tap changers, switched shunts, and remote generator voltage
regulation (Secs.~\ref{sec:remote}--\ref{sec:oltc}). Executed at solve time these
loops terminate on a best-effort basis, and their bus-traversal order depends on
the interpreter's hash seed, which made the converged profile vary between runs
(the minimum bus voltage ranged over \SIrange{0.82}{0.97}{pu} for an identical
command). We remove this nondeterminism by \emph{freezing} the operating point
once, offline, and distributing the solved snapshot---exactly as a saved PSS/E case
carries its solved taps and setpoints. A dedicated deterministic solver settles the
taps and reactive limits together (a relaxed bootstrap, then a coupled
reactive-limit loop that moves at most one tap step per pass, with a deadband and a
best-snapshot anti-hunt rule), and the result is baked into the CSV inputs: the
OLTC ratios into the transformer ratio fields, each generator's operating terminal
voltage into its setpoint with remote regulation collapsed to local control
(\texttt{reg\_bus}~$=0$), and the solved bus voltages and angles into the bus
table. With the control loops thus absorbed into the data, the published model
solves in a single deterministic Newton--Raphson pass---reactive-limit enforcement
and HVDC converter settling only---and reproduces the same operating point on every
run, with no tap or shunt control flags. The active controllers of
Secs.~\ref{sec:remote}--\ref{sec:oltc} are thus the optional online mode that this
one-time freezing replaces with a single deterministic settlement; re-running them
on the already-frozen data would double-regulate and must be avoided. One further
guard is applied after every data revision: because the slack absorbs any
generation--load imbalance as a free variable, an imbalance shows up in
\emph{no} power flow diagnostic; the settled slack output is therefore checked
explicitly against the slack unit's governor band, and the non-slack dispatch is
rebalanced whenever it drifts.

\subsection{Zero-impedance branch regularization}
Bus-coupler/section breakers appear as (near-)zero-impedance branches. Without the
node-merge that PSS/E applies internally, such branches make the admittance matrix
nearly singular and Newton--Raphson diverges.
A small lower bound ($z_{\min}=$\SI{e-4}{pu}) is imposed on such series
impedances, restoring conditioning with negligible effect on the solution.

\subsection{Element models}
Lines use the $\pi$ model: the series impedance $R+jX$ plus a single charging
susceptance $b^{\text{pu}}$ that the solver splits as $b^{\text{pu}}/2$ at each
end (converted to an equivalent capacitance). The optional end-shunt fields
($g/b$ from/to) are additive lumped shunts, set to zero here so that charging is
represented once. Two-winding transformers carry an off-nominal complex turns
ratio (no phase shift here); three-winding units pass winding-pair short-circuit
impedances to \texttt{pandapower}, which forms the star equivalent internally
(including a possibly negative middle leg for autotransformers), with the
nominal-voltage scaling pre-compensated. Loads support the full ZIP
(constant-impedance/constant-current/constant-power) form but are constant-power
here. Shunts are constant admittance; FACTS are zero-$P$ voltage-controlling
generators.

\section{Operating-Point Library}
\label{sec:oplib}

A single frozen snapshot cannot represent the range of conditions---overnight
valleys, the solar-driven midday minimum of net load, and the evening
peak---under which planning studies must hold. The dataset therefore ships,
alongside the reference case, a library of \num{24} hourly operating points for
a reference peak day (13 August 2025), each derived by the same deterministic
pipeline and frozen in the same single-pass form.

\paragraph{Derivation pipeline}
For each hour, a 48-hour security-constrained unit commitment (SCUC) is solved
as a mixed-integer linear program over the economic layer of
Sec.~\ref{sec:layers}: binary commitment with minimum up/down times, ramp
limits, configuration-based combined-cycle transitions, the five Korean reserve
products, and must-run renewables under hourly provincial profiles. Network
security is imposed on the DC approximation by \emph{lazy} constraint
generation: base-case and post-contingency flows are expressed through power
transfer and line outage distribution factors (PTDF/LODF), and violated $N-1$
(and selected same-corridor $N-2$) cuts are added iteratively until no
violation remains. The contingency set follows the Korean reliability code:
single circuits, EHV transformer banks, single generating units, and single
HVDC poles. The committed schedule is then dispatched, and the hour's snapshot
is reconciled in AC form---generator voltage schedules re-settled, reactive
adequacy checked (with additional units committed for voltage support where the
SCUC schedule alone cannot hold the profile), and the slack verified against
its governor band---before the operating point is frozen exactly as in
Sec.~\ref{sec:freeze}.

\paragraph{Outcome}
All \num{24} hourly cases solve as a single connected AC operating point with
reactive limits enforced. Across the library the transmission voltage profile
stays within the planning band at every hour, and successive data revisions have
eliminated all thermal overloads (the last residual, a Gwangju-area 154\,kV
corridor binding at the evening net-load peak, was resolved by a documented
circuit correction in the current release); the most heavily loaded circuits in
the early-evening hours now reach approximately \SI{99}{\percent} of their
continuous rating, so the library deliberately retains the system's thin
evening margins rather than masking them. The library exposes, on one
consistent network, the phenomena that distinguish modern Korean operations:
the evening net-load peak with zero solar output, midday renewable surplus with
reduced synchronous commitment, and overnight coal-set marginal prices
consistent with the published market record.

\section{The Reference Operating Point}
\label{sec:results}

\begin{figure}[htbp]
\centering
\includegraphics[width=\columnwidth]{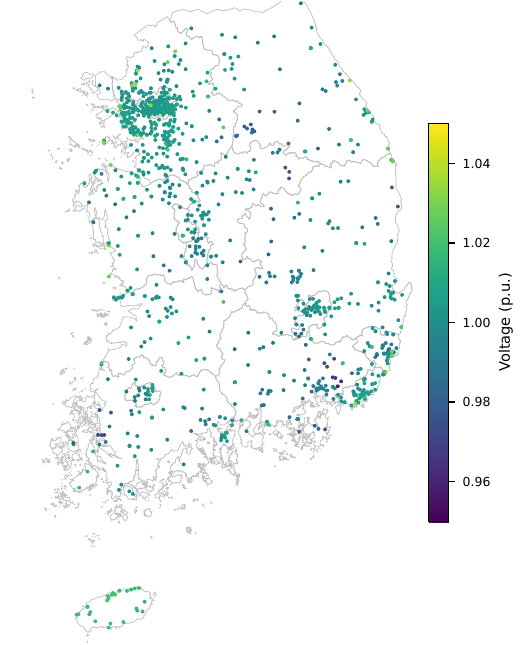}
\caption{Solved transmission bus-voltage profile at the evening-peak snapshot
(marker color, p.u.).}
\label{fig:pf}
\end{figure}

We evaluate the reference snapshot: the evening-peak hour of the reference
day, serving \SI{91.3}{\giga\watt}. Its demand level is the product of the
hourly shape of the EPSIS 2025 summer peak day (13 August 2025) and a national
anchor equal to the \SI{96}{\giga\watt} market-demand maximum recorded on
25 August 2025---a condition under which solar output is zero and
the conventional fleet carries the net-load maximum, which is where the Korean
system is now actually stressed. The
solution is a single connected network (no islands) and converges with reactive
limits enforced. Because the operating point is frozen into the data
(Sec.~\ref{sec:freeze}), the snapshot reproduces in a single deterministic
Newton--Raphson pass, identical on every run. Table~\ref{tab:pf} summarizes the
operating point and Fig.~\ref{fig:pf} maps the solved voltages.

\begin{table}[htbp]
\centering
\caption{Reference operating point (evening-peak snapshot).}
\label{tab:pf}
\small
\begin{tabular}{@{}lr@{}}
\toprule
Quantity & Value \\
\midrule
Total generation & \SI{93.4}{\giga\watt} \\
Total load & \SI{91.3}{\giga\watt} \\
Active-power losses & \SI{2.09}{\giga\watt} (\SI{2.3}{\percent}) \\
Mean transmission voltage & \SI{1.003}{pu} \\
Min / max transmission voltage & \SI{0.950}{pu} / \SI{1.035}{pu} \\
EHV/154\,kV buses below \SI{0.95}{pu} & 1 (at \SI{0.9498}{pu}) \\
Line circuits above \SI{100}{\percent} loading & 0 (max \SI{96.9}{\percent}) \\
HVDC transfer & \SI{1.29}{\giga\watt} over 4 links \\
Connected components & 1 (+ Jeju DC island) \\
Reactive limit enforcement & converged \\
Dangling bus references & 0 \\
\bottomrule
\end{tabular}
\end{table}

\paragraph{Voltage and flows}
Across the EHV and 154\,kV network the voltage profile (Fig.~\ref{fig:pf}) is
well within limits: the mean is \SI{1.003}{pu} and every transmission bus lies
within the $\pm$\SI{5}{\percent} planning band except a single 345\,kV bus in
southwestern Jeolla that grazes the lower limit at \SI{0.9498}{pu}---a
documented reactive-support margin discussed in Sec.~\ref{sec:limitations}. The
Seoul Metropolitan Area holds nominal voltage despite importing roughly
\SI{12}{\giga\watt}, and no line circuit exceeds its continuous rating (mean
loading \SI{19}{\percent}, maximum \SI{96.9}{\percent}). One 345/154\,kV
substation carries its banks marginally above rating even at the four-bank
maximum (\SI{108}{\percent}); the model thereby identifies an actual
transformer-capacity bottleneck rather than hiding it by unsupported
reinforcement.

\paragraph{Cross-check against KPG-193}
The independent public KPG-193 model~\cite{song2024kpg193}, built at far coarser
resolution, reports the same structural behavior our snapshot exhibits:
coastal-concentrated generation, a metropolitan load pocket drawing approximately
\SI{40}{\percent} of national demand, and the resulting long-distance northward
bulk transfer on the 345/765\,kV and HVDC corridors. Our solved operating point
is consistent with this picture; because the two models differ greatly in
resolution and neither is tuned to the other, we claim structural agreement rather
than a bus-for-bus match.

\section{Model Validation and Consistency}
\label{sec:validation}

Because no public full-resolution Korean case exists to compare against
bus-for-bus, we validate the model against \emph{public aggregate benchmarks} and
through internal physical consistency. Four checks are reported.

\begin{figure*}[t]
\centering
\includegraphics[width=0.78\textwidth]{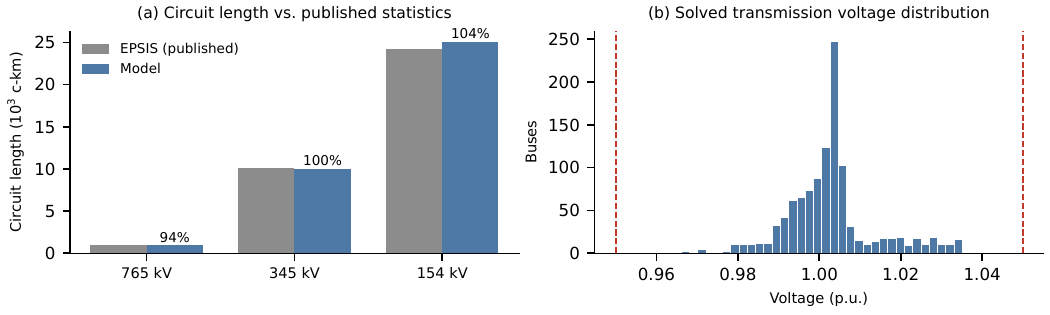}
\caption{Public-benchmark validation. (a) Assembled circuit length versus
published EPSIS, by voltage level. (b) Solved transmission bus-voltage
distribution at the evening-peak snapshot.}
\label{fig:validation}
\end{figure*}

\paragraph{Transmission length vs EPSIS}
The most direct quantitative anchor is total transmission circuit length by
voltage level, which EPSIS publishes~\cite{epsis}. The assembled network reaches
94\%, 100\%, and 104\% of the published 765, 345, and 154\,kV circuit-km
respectively (Fig.~\ref{fig:validation}a), and $\approx$102\% in aggregate. Since
the 345 and 154\,kV layout is taken from real OSM geometry and only the
unreachable substations are gap-filled, this agreement is a check on the
\emph{amount} of transmission rather than a fitted result.

\paragraph{Facility inventories vs EPSIS}
The same statistics anchor the equipment populations: the model carries the
EPSIS-exact eight 765\,kV substations, \num{799} 154\,kV distribution stations
against a published \num{788}, 345/154\,kV transformer capacity within
\SI{3}{\percent} and 765/345\,kV capacity within \SI{4}{\percent} of the
published figures (the 154/22.9\,kV firm-bank population lands at
\SI{110}{\percent}, the deliberate overshoot of an $N$$-$$1$ firm rule applied
at peak load), and aggregate capacitor/reactor capacities at
99\%/101\% of the published reactive-equipment totals. The generation fleet is
reconciled unit-by-unit against the public registered-generator inventory, and
the provincial allocation of demand deviates from the published shares by
1.9 percentage points summed over all 17 provinces. Where a gap remains
(the model resolves 106 of the 122 published 345\,kV substations), it is because
the missing stations are absent from all public geographic sources, and we
decline to invent them beyond the documented synthetic procedure.

\paragraph{Voltage profile and convergence}
The AC power flow converges to a single connected operating point with generator
reactive limits enforced, and the solved transmission voltages form a tight,
well-behaved distribution around \SI{1.003}{pu} (Fig.~\ref{fig:validation}b).
Convergence under enforced reactive limits---rather
than only with limits relaxed---is itself evidence that the reactive resources
and dispatch are physically self-consistent under peak load; the same holds at
every hour of the operating-point library (Sec.~\ref{sec:oplib}).

\paragraph{Cross-check against KPG-193}
The only other public Korean synthetic model, KPG-193~\cite{song2024kpg193}, is
built at far coarser resolution (193 buses) and is exercised by AC optimal power
flow rather than a fixed power flow snapshot, so a bus-for-bus voltage comparison
is not meaningful. Qualitatively, however, both models show the same behavior:
coastal-concentrated generation, a metropolitan area drawing approximately
\SI{40}{\percent} of national demand, long-distance northward bulk transfer on the
345/765\,kV and HVDC corridors, and voltage stress concentrated on the
weakly-meshed periphery---which is the consistency we can legitimately claim across
resolutions. The underlying solver is the standard \texttt{pandapower}
Newton--Raphson formulation~\cite{thurner2018pandapower}, whose correctness is
established independently in the literature; here we add the control loops of
Sec.~\ref{sec:solver} on top of it.

\section{Limitations}
\label{sec:limitations}

We state the model's limitations explicitly. They fall into two groups: quantities
that could not be obtained from public data and were therefore estimated, and
modeling simplifications inherent to snapshot-based representation.

\medskip
\noindent\emph{Quantities estimated because they are not in public data.}
\begin{enumerate}[leftmargin=1.3em,itemsep=0.2ex,topsep=0.3ex]
\item \textbf{Shunt and FACTS reactive set points.} The per-device set points of
shunt voltage-compensation devices (switched capacitors/reactors) cannot be
determined from public data. They are placed by engineering heuristics at
electrically weak buses and sized to the operating snapshot; locations, counts,
and \emph{aggregate} capacity are reconciled against public statistics, but
individual device ratings are approximate.
\item \textbf{On-load tap-changer settings.} OLTC tap positions and regulated
voltage set points are likewise unavailable publicly. Taps are frozen at
plausible snapshot values by default (Sec.~\ref{sec:oltc}); the optional active
mode settles them by a control heuristic rather than from recorded operator
settings.
\item \textbf{Partly estimated topology.} While the 345 and 154\,kV layout is
taken from real OSM geometry where coverage exists, the remaining
unreachable substations are connected
by the geographic MST heuristic of Sec.~\ref{sec:topology}, so those corridors are
estimated rather than observed; about \num{90} substations remain at synthetic
(golden-angle) positions. Per-circuit counts are feeder-tag estimates
(aggregate circuit-km matches EPSIS, but individual circuits are approximate).
\item \textbf{Electrical parameters and identifiers.} Line $R/X/B$ use standard
per-kilometer values for Korean overhead conductors~\cite{song2024kpg193} and
XLPE cables rather than measured per-circuit impedances; zero-sequence data and
neutral grounding are typical-value seeds (Sec.~\ref{sec:sequence}). Substation
coordinates are crowd-sourced ($\pm$ hundreds of metres), and bus numbers and
names are synthetic.
\item \textbf{Standard-parameter companion layers.} The dynamic, protection, and
economic layers of Sec.~\ref{sec:layers} carry published typical parameters,
scaled by rating---not plant-specific tuning or utility setting files---and
should be treated accordingly in any study whose conclusions depend on them.
\end{enumerate}

\medskip
\noindent\emph{Modeling simplifications and residual stress.}
\begin{enumerate}[leftmargin=1.3em,itemsep=0.2ex,topsep=0.3ex,resume]
\item \textbf{Documented residual weaknesses.} One 345\,kV bus in southwestern
Jeolla grazes the \SI{0.95}{pu} limit at the evening peak, and one 345/154\,kV
substation loads its banks to $\approx$\SI{108}{\percent} even at the four-bank
maximum. Both correspond to areas where the actual system concentrates
reinforcement (planned synchronous condensers and substation expansion); since
only in-operation facilities are modeled, we record these as findings rather
than patch them with unsupported equipment. Radial island feeders near the Jindo
HVDC converter sit lowest among distribution buses for the same reason.
\item \textbf{Snapshot-based representation.} The model represents frozen
operating points---one reference snapshot plus the 24-hour library of
Sec.~\ref{sec:oplib}---with constant-power loads; it is not a continuous time
series, and enforcing generator reactive limits at peak remains near the
boundary of what a single power flow (versus a coordinated reactive optimal
power flow) can achieve.
\item \textbf{Mixed data vintages.} The public inputs are taken from the most
recent editions available at construction and do \emph{not} share a single
reference year: the EPSIS statistics used for calibration are 2024--2025
editions, the fleet status and demand anchor are August 2025, and the
OpenStreetMap geometry was retrieved in 2026. The model is therefore best read
as a representative composite of the recent Korean system rather than a
reconstruction of its state at any single instant.
\end{enumerate}

\section{Conclusion}
\label{sec:conclusion}

We have presented the GIST Korea test system, a geographically explicit
synthetic model of the Korean power grid built \emph{entirely} from public data
and exercised by AC power flow. Its distinguishing feature is that the 345 and
154\,kV transmission graph is derived from the real OpenStreetMap power layer by
a multi-source shortest-path reassembly with feeder-based circuit counting and
explicit overhead/underground section modeling, with only
unreachable substations gap-filled by a geographic MST, and is calibrated to
published circuit-length and facility statistics. The model is distributed as
frozen operating points---tap ratios, generator setpoints, and bus voltages
settled once offline and baked into the data---so a single deterministic
Newton--Raphson pass reproduces the evening-peak snapshot on every run, at an
operating point whose transmission voltage profile is within operating limits
and consistent with the independent public KPG-193 model; a
security-constrained unit-commitment pipeline extends this to a 24-hour library
of operating points on the same network. Companion layers of standard-parameter
dynamic, protection, and economic data make the case exercisable well beyond
power flow. Distributed as a CSV dataset that follows PSS/E conventions, with
reproducible build tooling, the model is intended as a citable platform,
assembled entirely from public data, for power flow, contingency,
reactive-planning, and decarbonization studies of one of the world's most
structurally distinctive grids.

The model is a \emph{living} dataset: it is under continuous refinement, and
component counts---including the number of buses---change between releases as
synthetic elements are re-identified with real facilities and public sources
improve. The description in this paper corresponds to release v5.2.0
(July 2026); the current release, its changelog, and the interactive map are
maintained on the GIST Power System Laboratory web
page~\cite{gist_dataset,gist_map}. Future work includes future-year scenario
models built from the national supply plans, reactive optimization under
enforced limits, and automatic $N$$-$$1$ contingency screening.

\section*{Data Availability}
The dataset (\texttt{bus}, \texttt{gen}, \texttt{line}, \texttt{load},
\texttt{shunt}, \texttt{trafo2w}, \texttt{trafo3w}, \texttt{facts},
\texttt{dcline}, \texttt{vscdcline} CSV files, together with the
\texttt{dynamics}, \texttt{protection}, and economic data layers and the
operating-point library), the topology and power flow maps, and the build and
power flow scripts are available from the GIST Power System Laboratory at
\url{https://psl.gist.ac.kr/prog/bbsArticle/BBSMSTR_000000012527/list.do}~\cite{gist_dataset}.
An interactive, browser-based rendering of the network topology is also available
online~\cite{gist_map}. Because the model is continuously updated, readers
should consult the laboratory page for the current release and its changelog.
All of it is derived solely from publicly available
sources; no confidential or operator-internal data is included. The schema is
summarized in the Appendix (Table~\ref{tab:schema}).

\section*{Acknowledgment}
This work relies on the publicly available OpenStreetMap/OpenInfraMap data and
the public statistics of KPX/EPSIS and the Korea Energy Agency, whose maintainers
we gratefully acknowledge.

\appendix
\section{CSV Input Schema}
\label{app:schema}
Table~\ref{tab:schema} lists the columns of each core CSV input file; the data
conventions are noted below the table.
\begin{table*}[!t]
\centering
\caption{Columns of each core CSV input file.}
\label{tab:schema}
\small
\setlength{\emergencystretch}{3em}
\begin{tabular}{@{}l p{0.82\textwidth}@{}}
\toprule
File & Columns \\
\midrule
\texttt{bus.csv} & \texttt{bus, name, region, base\_kv, volt\_class, type, area, area\_name, zone, vm\_pu, va\_deg, vmax\_pu, vmin\_pu, in\_service, lat, lon} \\
\texttt{gen.csv} & \texttt{bus, id, name, p\_mw, q\_mvar, vset\_pu, q\_max\_mvar, q\_min\_mvar, p\_max\_mw, p\_min\_mw, mbase\_mva, reg\_bus, is\_slack, in\_service, qctrl, qv\_*} \\
\texttt{load.csv} & \texttt{bus, id, name, p\_mw, q\_mvar, const\_i\_*, const\_z\_*, motor\_pct, conv\_pct, area, zone, in\_service} \\
\texttt{shunt.csv} & \texttt{bus, id, name, kind, p\_mw, q\_mvar, in\_service} \\
\texttt{line.csv} & \texttt{from\_bus, to\_bus, id, name, r\_pu, x\_pu, b\_pu, g/b\_from/to\_pu, rate\_a/b/c\_mva, length\_km, in\_service, cable\_km, r0\_pu, x0\_pu, b0\_pu} \\
\texttt{trafo2w.csv} & \texttt{hv\_bus, lv\_bus, r\_pu, x\_pu, b\_pu, ratio, tap\_*, shift\_deg, vn\_hv/lv\_kv, sn\_mva, rate\_*\_mva, cont\_bus, in\_service, conn, zn\_hv/lv\_r/x\_ohm} \\
\texttt{trafo3w.csv} & \texttt{hv/mv/lv\_bus, r/x\_(hv\_mv,mv\_lv,lv\_hv)\_pu, ratio\_hv/mv/lv, tap\_*, vn\_hv/mv/lv\_kv, rate\_*\_mva, cont\_bus, in\_service, conn, zn\_hv/mv/lv\_r/x\_ohm} \\
\texttt{facts.csv} & \texttt{name, bus, bus\_name, mode, vset\_pu, shmax\_mvar, pdes\_mw, qdes\_mvar} \\
\texttt{dcline.csv} & per-pole LCC converter records: DC schedule and resistance, bridge counts, firing/extinction angle bands, commutating reactances, converter transformer taps, filter Mvar \\
\texttt{vscdcline.csv} & VSC converter records: per-terminal control mode (DC power/voltage, AC voltage/Q), converter loss coefficients, MVA and reactive limits \\
\bottomrule
\end{tabular}
\par\vspace{3pt}
{\scriptsize\raggedright The schema follows PSS/E data conventions. Slash groups enumerate
variants (e.g., \texttt{rate\_a/b/c\_mva}) and \texttt{*} a suffix family (e.g.,
\texttt{const\_i\_*}~$=$~\texttt{const\_i\_p\_mw}, \texttt{const\_i\_q\_mvar}). All
impedances/admittances are per-unit on a \SI{100}{\mega\voltampere} base;
transformer turns ratios are off-nominal per-unit; \texttt{cable\_km} is the
underground length of the circuit and \texttt{r0/x0/b0} its zero-sequence
parameters; \texttt{conn} and \texttt{zn\_*} encode winding connection and
neutral grounding. The \texttt{dynamics/}, \texttt{protection/}, and
\texttt{opt/} sub-directories carry the standard-parameter layers of
Sec.~\ref{sec:layers}.\par}
\end{table*}



\begin{thebibliography}{99}
\bibitem{birchfield2017structural}
A.~B. Birchfield, T. Xu, K.~M. Gegner, K.~S. Shetye, and T.~J. Overbye, ``Grid
structural characteristics as validation criteria for synthetic networks,''
\emph{IEEE Trans. Power Syst.}, vol.~32, no.~4, pp.~3258--3265, 2017.

\bibitem{gegner2016methodology}
K.~M. Gegner, A.~B. Birchfield, T. Xu, K.~S. Shetye, and T.~J. Overbye, ``A
methodology for the creation of geographically realistic synthetic power flow
models,'' in \emph{Proc. IEEE Power Energy Conf. Illinois (PECI)}, 2016, pp.~1--6.

\bibitem{birchfield2017validation}
A.~B. Birchfield, E. Schweitzer, H. Athari, T. Xu, T.~J. Overbye, A. Scaglione,
and Z. Wang, ``A metric-based validation process to assess the realism of
synthetic power grids,'' \emph{Energies}, vol.~10, no.~8, p.~1233, 2017.

\bibitem{birchfield2018convergence}
A.~B. Birchfield, T. Xu, and T.~J. Overbye, ``Power flow convergence and reactive
power planning in the creation of large synthetic grids,'' \emph{IEEE Trans.
Power Syst.}, vol.~33, no.~6, pp.~6667--6674, 2018.

\bibitem{xu2017economic}
T. Xu, A.~B. Birchfield, K.~M. Gegner, K.~S. Shetye, and T.~J. Overbye,
``Application of large-scale synthetic power system models for energy economic
studies,'' in \emph{Proc. 50th Hawaii Int. Conf. Syst. Sci. (HICSS)}, 2017.

\bibitem{mateo2024td}
C. Mateo \emph{et al.}, ``Building and validating a large-scale combined
transmission and distribution synthetic electricity system of Texas,''
\emph{Int. J. Electr. Power Energy Syst.}, vol.~159, Art.~no.~110037, 2024.

\bibitem{song2024kpg193}
G. Song and J. Kim, ``KPG~193: A synthetic Korean power grid test system for
decarbonization studies,'' \emph{arXiv:2411.14756}, 2024.

\bibitem{thurner2018pandapower}
L. Thurner \emph{et al.}, ``pandapower---An open-source Python tool for convenient
modeling, analysis, and optimization of electric power systems,'' \emph{IEEE
Trans. Power Syst.}, vol.~33, no.~6, pp.~6510--6521, 2018.

\bibitem{zimmerman2011matpower}
R.~D. Zimmerman, C.~E. Murillo-S\'anchez, and R.~J. Thomas, ``MATPOWER:
Steady-state operations, planning, and analysis tools for power systems research
and education,'' \emph{IEEE Trans. Power Syst.}, vol.~26, no.~1, pp.~12--19, 2011.

\bibitem{psse2013}
Siemens PTI, \emph{PSS/E Program Operation Manual,
v33}, 2013.

\bibitem{grainger1994}
J.~J. Grainger and W.~D. Stevenson, \emph{Power System Analysis}. New York:
McGraw-Hill, 1994.

\bibitem{kundur1994}
P. Kundur, \emph{Power System Stability and Control}. New York: McGraw-Hill, 1994.

\bibitem{osm}
OpenStreetMap contributors, ``Planet dump,'' \url{https://www.openstreetmap.org};
power layer via OpenInfraMap, \url{https://openinframap.org}, accessed 2026.

\bibitem{epsis}
Korea Power Exchange (KPX), ``Electric Power Statistics Information System
(EPSIS),'' \url{https://epsis.kpx.or.kr}, accessed 2026.

\bibitem{kea}
Korea Energy Agency (KEA), ``New and Renewable Energy statistics / regional
deployment data,'' \url{https://www.energy.or.kr}, accessed 2026.

\bibitem{kesis}
Korea Energy Economics Institute, ``Korea Energy Statistical Information System
(KESIS): regional electricity consumption,'' \url{https://www.kesis.net},
accessed 2026.

\bibitem{kepcostat}
Korea Electric Power Corporation (KEPCO), ``Statistics of Electric Power in
Korea,'' annual report, \url{https://home.kepco.co.kr}, accessed 2026.

\bibitem{jeju_diagram}
T. Lee and Y. Lee, ``Expansion of renewable energy in Jeju and directions for
stable power-system operation (in Korean),'' \emph{Energy Focus}, Korea Energy
Economics Institute (KEEI), Winter 2020, pp.~48--63, Fig.~5 (Jeju power-system
one-line diagram).

\bibitem{sinnott1984}
R. W. Sinnott, ``Virtues of the haversine,'' \emph{Sky and Telescope}, vol.~68,
no.~2, p.~159, 1984.

\bibitem{cover1967}
T. Cover and P. Hart, ``Nearest neighbor pattern classification,''
\emph{IEEE Trans. Inf. Theory}, vol.~13, no.~1, pp.~21--27, 1967.

\bibitem{prim1957}
R. C. Prim, ``Shortest connection networks and some generalizations,''
\emph{Bell Syst. Tech. J.}, vol.~36, no.~6, pp.~1389--1401, 1957.

\bibitem{dijkstra1959}
E. W. Dijkstra, ``A note on two problems in connexion with graphs,''
\emph{Numer. Math.}, vol.~1, pp.~269--271, 1959.

\bibitem{kostat_admin}
Statistics Korea (KOSTAT), administrative-district boundary data, distributed via
the \emph{southkorea-maps} dataset, \url{https://github.com/southkorea/southkorea-maps},
accessed 2026.

\bibitem{gist_map}
Y.-S. Kim, ``The GIST Korea test system: interactive grid map,''
\url{https://yskimpsl.github.io/GIST-2064-bus-test-system/grid_map_reconstructed.html},
accessed 2026.

\bibitem{tinney1967}
W. F. Tinney and C. E. Hart, ``Power flow solution by Newton's method,''
\emph{IEEE Trans. Power App. Syst.}, vol.~PAS-86, no.~11, pp.~1449--1460,
Nov. 1967.

\bibitem{gist_dataset}
Y.-S. Kim, ``The GIST Korea test system: Korean power-grid dataset,'' GIST
Power System Laboratory,
\url{https://psl.gist.ac.kr/prog/bbsArticle/BBSMSTR_000000012527/list.do},
accessed 2026.

\bibitem{huangfu2018}
Q. Huangfu and J.~A.~J. Hall, ``Parallelizing the dual revised simplex method,''
\emph{Math. Program. Comput.}, vol.~10, no.~1, pp.~119--142, 2018.

\bibitem{cui2021andes}
H. Cui, F. Li, and K. Tomsovic, ``Hybrid symbolic--numeric framework for power
system modeling and analysis,'' \emph{IEEE Trans. Power Syst.}, vol.~36, no.~2,
pp.~1373--1384, 2021.

\bibitem{wecc2014}
WECC Renewable Energy Modeling Task Force, ``WECC second generation renewable
energy system models,'' Western Electricity Coordinating Council, Tech. Rep.,
2014.

\bibitem{ieee_c37_102}
\emph{IEEE Guide for AC Generator Protection}, IEEE Standard C37.102, 2006.

\bibitem{ieee_c37_106}
\emph{IEEE Guide for Abnormal Frequency Protection for Power Generating Plants},
IEEE Standard C37.106, 2003.

\bibitem{iec60255}
\emph{Measuring Relays and Protection Equipment---Part 151: Functional
Requirements for Over/Under Current Protection}, IEC Standard 60255-151, 2009.
\end{thebibliography}
\end{document}